\documentclass[prd,preprintnumbers,superscriptaddress,nofootinbib,floatfix,twocolumn,notitlepage]{revtex4}
\usepackage[dvips]{graphicx}
\usepackage{dcolumn} 
\usepackage{bm}
\usepackage{epsfig,amsmath,amssymb,verbatim,mathrsfs,array,layout,textcomp,amssymb,latexsym,slashed}
\usepackage{color}
\usepackage{hyperref}
\usepackage[utf8]{inputenc}

\input epsf.tex
\newcommand{\beq}{\begin{eqnarray}}
\newcommand{\eeq}{\end{eqnarray}}

\def\ltap{\ \raise.3ex\hbox{$<$\kern-.75em\lower1ex\hbox{$\sim$}}\ }
\def\gtap{\ \raise.3ex\hbox{$>$\kern-.75em\lower1ex\hbox{$\sim$}}\ }

\def\be{\begin{equation}}
\def\ee{\end{equation}}
\def\bea{\begin{eqnarray}}
\def\eea{\end{eqnarray}}

\pdfoutput=1

\newcommand{\cO}{\mathcal{O}}
\newcommand{\cL}{\mathcal{L}}

\definecolor{newred}{rgb}{0.5,0.1,0}
\definecolor{darkgreen}{rgb}{0.0,0.7,0.2}

\begin{document} 

\title{Probing New Physics with Isotope Shift Spectroscopy}

\author{C\'edric Delaunay}
\email{cedric.delaunay@lapth.cnrs.fr}
\affiliation{LAPTh, Universit\'e Savoie Mont Blanc, CNRS B.P. 110, F-74941 Annecy-le-Vieux, France}
\author{Yotam Soreq}
\email{soreqy@mit.edu}
\affiliation{Center for Theoretical Physics, Massachusetts Institute of Technology, Cambridge, MA  02139, U.S.A.}

\begin{flushright}
\preprint{\scriptsize LAPTH-007/16\vspace*{.1cm}}
\preprint{\scriptsize  MIT-CTP/4769\vspace*{.1cm}}
\end{flushright}
\vskip .025in

\begin{abstract}
We investigate the potential to probe physics beyond the Standard Model with isotope shift measurements of optical atomic clock transitions. We first derive the reach for generic new physics above the GeV scale at the effective field theory level, as well as estimate the limits on possible new spin-independent forces mediated by sub-GeV states coupled to electrons and neutrons. We also study the weak force and show that isotope shifts could provide strong constraints on the $Z^0$ couplings to valence quarks, which complement precision observables at LEP and atomic parity violation experiments. Finally, motivated by recent experimental hints of a new 750\,GeV resonance in diphotons, we also consider the potential to probe its parity-preserving couplings to  electrons, quarks and gluons with this method. In particular,  combining the diphoton signal with indirect constraints from $g_e-2$ and isotope shifts in Ytterbium allows to probe the resonance coupling to electrons with unprecedented precision.
\end{abstract}

\maketitle

\section{Introduction} \label{sec:intro}

The Standard Model of particle physics~(SM) describes, with a huge success, a vast amount of phenomena in a wide range of energy scales. With the discovery of a $125\,$GeV Higgs boson~\cite{CMSdisco,ATLASdisco}, the SM is a theoretically consistent theory up to scales much larger than the electroweak scale. However, it is well establish that the SM is not a complete description of Nature. For example, it can not account for the observed  matter-antimatter asymmetry  of our Universe,  nor neutrino oscillations and it does not contain a viable dark matter candidate. All these observations require physics beyond the SM, although none of them points to a specific mass scale. 

Much of the effort in searching for new physics is done at collider experiments, like ATLAS and CMS at the LHC, which stand at the energy frontier and with $B$-factories and LHCb at the intensity frontier. A~third, complementary approach relies on high precision measurements of clean (both theoretically and experimentally) low energy processes. Prominent examples are the electric and magnetic dipole moments of the electron~\cite{Baron:2013eja,PhysRevLett.100.120801,bouchendira2011new}. Atomic parity violating~(APV) amplitudes in heavy atoms~\cite{wood1997measurement,guena2005measurement} also supply valuable observables. Recently, Ref.~\cite{Delaunay:2016brc} proposed to use precision isotope shift (IS) measurements in atomic clock transitions. Although the method was originaly proposed to probe the Higgs couplings to the electron and the first generation quarks, it is in principle sensitive to many other phenomena with interactions to electrons and nuclei, as long as those are not proportional to the electric charge. A noteworthy example is the weak force whose parity-conserving part could lead to measurable effects, even for couplings of the $Z^0$ boson at their SM values~\cite{Delaunay:2016brc}. This letter aims at evaluating the possibility to probe deviations of the $Z^0$ coupling to SM fermions as well as other forms of new physics around and above the GeV scale with state-of-the-art IS measurements.\\

\section{Isotope Shifts in atomic clocks}\label{sec:IS}

Atomic clocks involve very narrow optical transitions in heavy atoms and ions~\cite{2014arXiv1401.2378P}. Frequency comparisons of such transitions are already extremely precise in several systems~\cite{rosenband2008frequency,Innsbruck,PTB,NRC,JunYe}, with a relative accuracy down to $10^{-18}$~\cite{JunYe}. Frenquency shifts between isotopes in some of them are also measured with very good accuracy, see {\it e.g.}~\cite{ChiaveriniClockIS}. 
The frequency shift between two isotopes of mass $A$ and $A'$ for a transition $i$ is predicted to be~\cite{ISKing,Delaunay:2016brc}
\beq\label{IStheory}
\delta\nu^i_{AA'} = K_i \mu_{AA'}+F_i\delta\langle r^2\rangle_{AA'} + X^i_{AA'},
\eeq 
where $\mu_{AA'}\equiv m_A^{-1}-m_{A'}^{-1}$ is the electron-nucleus reduced mass change, $\delta \langle r^2\rangle_{AA'}\equiv \langle r^2_A\rangle-\langle r^2_{A'}\rangle$ is the difference in the nuclear charge distribution variance, while $K_i$ and $F_i$ are electronic constants depending only on the transition, not on the nuclear parameters. The first term on the RHS is due to nuclear mass shift~(MS), the second term denotes the so-called volume or field shift~(FS) and $X^i_{AA'}$ generically represents possible new physics contributions. The $K_i$ and $F_i$ constants and $\delta\langle r^2\rangle_{AA'}$ are challenging to calculate accurately from first principles.  Therefore, the extraction of $X_{AA'}^i$ from an IS measurement in a single transition $i$ is by far limited by theory uncertainties. Nevertheless, theory uncertainties largely cancel out when IS measurements in two distinct transitions (with the same isotopes) are plotted against each other in a so-called King plot~\cite{King:63}.
 IS measurements then offer a way to extract (or bound) {\it differences } of $X_{AA'}^i$ contributions which is potentially only limited by experimental errors. This requires notably that non-linear MS and FS corrections to the King plot are sufficiently suppressed. Although this is expected on theoretical grounds~\cite{Palmer,blundell1987reformulation,Delaunay:2016brc}, confirmation from many-body simulations with relativistic corrections remains necessary. As we show in the following sections, this method then turn IS measurements into a probe of many scenarios for physics beyond the SM with unprecependented precision.

For sake of concreteness, we apply the above strategy using the two known narrow transitions in singly ionized Ytterbium Yb$^+$ ($Z=70$, $A=168,\dots,176$), namely the quadrupole (E2) and octupole (E3) electric transitions at 436$\,$nm and 467$\,$nm, respectively. Both of them involve a $6S$ orbital which is most sensitive to short range interactions through its large overlap with nuclei. Experiments have already demonstrated the ability to probe both transitions with an uncertainty below the Hz level~\cite{Godun:2014naa,PTB,huntemann2014improved} with good prospect for improvement~\cite{2016arXiv160203908H} and IS measurements with Hz precision~\cite{Nisbet-Jones}.

Finally, note that a critical point of the above approach is the validity of the factorized form of the MS and FS terms in Eq.~\eqref{IStheory} which is expected to break at some level, yielding non-linear King plots. However, observing that IS contributions are controlled by small parameters, namely $m_e \mu_{AA'}$ and $\delta \langle r^2\rangle_{AA'}/a_0^2$ where $a_0=(\alpha m_e)^{-1}$ is the Bohr radius, such non-linearities are expected to be at most an order of magnitude below the uncertainty currently limiting Yb$^+$ experiments~\cite{Delaunay:2016brc,futurework}. We therefore assume linear King plots in Yb$^+$ with Hz-level accuracy in this letter. In the case that,  in a near future, experimental uncertainties were reduced so to reveal MS/FS non-linearities, a series of dispositions could still be implemented to  keep improving their sensitivity to new physics. For instance, these effects could presumably be computed with sufficient accuracy in order to substract them from IS measurements and construct a King plot with the residuals. Another possibility, if the dependence of MS/FS non-linearities with the isotope mass difference differs from that of new physics, would be to use similar measurements with additional isotopes to calibrate them and isolate possible more fundamental contributions. 

\section{Effective Theory Analysis} \label{sec:EFT}

Atomic IS are modified by short distance physics which couples to electrons as well as quarks and/or gluons. We focus in this work on spin-independent interactions whose contributions to energy level shifts are enhanced by the number of nucleons. 
Therefore, the Lagrangian at the scale $\mu\approx\,$GeV, relevant for our effective field theory~(EFT) analysis is
\beq
	\cL_{\rm eff} = \cL_{\rm SM} + \frac{1}{\Lambda^2} \left[\sum_{q} c_{eq}^S \cO_{eq}^S + c_{eq}^V \cO_{eq}^V\right] + \frac{c_{eg}}{\Lambda^3}\cO_{eg} \, ,
\eeq
where $\cL_{\rm SM}$ is the Standard Model Lagrangian, and the higher-dimensional operators  
are
\beq
\cO_{eq}^{S} &=& (\bar e e)(\bar q q)\,,\label{OpeqS}\\
\cO_{eq}^{V} &=& (\bar e \gamma_\mu e)(\bar q\gamma^\mu q)\,,\label{OpeqV}\\
\cO_{eg}&=& \alpha_s (\bar e e)G_{\mu\nu}^aG^{\mu\nu\,a}\,,\label{Opeg}
\eeq
with $q=u,d,s$ in $\cO_{eq}^S$, and $q=u,d$ in $\cO_{eq}^V$. Note that, since vector currents are conserved, nucleon couplings are only sensitive to the vector currents of valence quarks. 
Since $Q=c,b,t$ quarks are integrated out at $\mu\approx\,$GeV, similar four-fermion operators of the form 
\beq
\cO_{eQ}^S=(\bar e e)(\bar QQ)\label{OpeQS}
\eeq
 only contribute as threshold corrections to $\cO_{eg}$, shifting its associated Wilson coefficient~(WC) such that
\beq\label{thresholdcorr}
 c_{eg}= \delta c_{eg}-\frac{1}{12\pi }\sum_{Q=c,b,t}c_{eQ}^S\frac{\Lambda}{m_Q}\,,
\eeq
where $\delta c_{eg}$ denotes genuine new physics contributions to $c_{eg}$. 

Within atomic systems of mass $A$ and atomic number $Z$, the above operators induce a local potential between nuclei and their bound electrons of 
\beq\label{EFTpotential}
V_{\rm EFT}(r)= -\frac{y_{eA}}{4\pi\Lambda^2}\frac{\delta(r)}{r^2}\,,
\eeq
where $y_{eA}\equiv y_{ep} Z+ y_{en} (A-Z)$. IS measurements can only probe the neutron contribution which relates to the above WC's (evaluated at $\mu=\Lambda$, with $\Lambda=1\,$TeV for definiteness) through~\cite{Shifman:1978zn,Belanger:2008sj,Xing:2011aa,Junnarkar:2013ac,microOmega} 
\beq\label{EFTcoupling}
y_{en}&\approx& 8.8 \,c_{eu}^S+11 \,c_{ed}^S+0.86 \,c_{es}^S-2.4\times 10^{-3}\,c_{eg}\nonumber\\
&&+c_{eu}^V+2c_{ed}^V\,,
\eeq
where $c_{eg}\approx \delta c_{eg}-49 \,c_{ec}^S-11\, c_{eb}^S - 0.17 \,c_{et}^S$. The resulting new physics contribution to the IS in Eq.~\eqref{IStheory} for $i=nS\to n'D$ or $n'F$ transitions is
\beq
X_{AA'}^i\big|_{\rm EFT} = 4 \,{\rm Hz}\times y_{en}(A-A')\frac{|\psi(0)|^2}{4a_0^{-3}}\left(\frac{{\rm TeV}}{\Lambda}\right)^2,
\eeq 
where $|\psi(0)|^2$ is the $nS$ electron density at the nucleus, which in the non-relativistic limit is approximately $\simeq 4.2Z(1+n_e)^2/(n a_0)^3$~\cite{AtomicBook,Delaunay:2016brc}. Following Ref.~\cite{Delaunay:2016brc}, the observation of linearity in a King plot constructed from IS measurements for the two Yb$^+$ clock transitions down to an accuracy of $\Delta \sim \cO($Hz$)$ would yield
\beq
|y_{en}|\lesssim \frac{0.02}{|1-F_{21}|}\left(\frac{\Lambda}{{\rm TeV}}\right)^2\left(\frac{\Delta}{{\rm Hz}}\right)\left(\frac{8}{A-A'}\right)\,.
\eeq
where $F_{21}\equiv F_2/F_1$. From this bound we derive in Table~\ref{tab:EFT} the reach  of each higher dimensional operators in Eqs.~\eqref{OpeqS}--\eqref{OpeQS}. The obtained limits for vector operators are stronger than those obtained from LEP2 measurements~\cite{Schael:2013ita} by a factor $2-3$, as well as comparable or slightly stronger than those from the $8\,$TeV and $13\,$TeV LHC~\cite{deBlas:2013qqa, ATLAS-CONF-2015-070}. For scalar operators, the above limits are stronger than LEP\,2 (whenever relevant) by about an order of magnitude, mostly due to the larger nuclear form factors.
\begin{table}[!t]
\begin{center}
\renewcommand{\arraystretch}{1.3}
\begin{tabular}{|c|c|c|}
\hline
operator  & Upper bound on $|c_i|$   &  Lower bound on $\Lambda_i$ [TeV] \\
$\cO_i$ & ($\Lambda=1\,$TeV) &   ($c=1$)\\
 \hline\hline
  $\cO^V_{eu}$ & $2.3\times10^{-2}$ & 6.6 \\
  $\cO^V_{ed}$ & $1.1\times10^{-2}$ & 9.3 \\
  \hline
  $\cO^S_{eu}$ & $2.6\times10^{-3}$ & 20 \\
  $\cO^S_{ed}$ & $2.1\times10^{-3}$ & 22 \\ 
  $\cO^S_{es}$ & $2.7\times10^{-2}$ & 6.1 \\ 
  \hline
  $\cO^S_{ec}$ & $0.20$& 2.3 \\
  $\cO^S_{eb}$ & $0.87$ & 1.1 \\ 
  $\cO^S_{et}$ & $56$ & 0.13\\  
  \hline
  $\cO_{eg}$ & $9.6$ & 0.47 \\  
  \hline
  \hline
\end{tabular}
\end{center}
\caption{Projected bounds of generic heavy new physics from isotope shift measurements in Yb$^+$ with an accuracy of $\Delta=\,1$Hz and assuming $A'-A=8$. The first column is the local operator of interest, while the second and third columns are the corresponding bounds on its Wilson coefficient for $\Lambda=1\,$TeV and its  effective scale for a coefficient of unity.}
\label{tab:EFT}
\end{table}%

\section{Parity-Conserving Weak Force} \label{sec:weak}

Consider now the weak force between nuclei and electrons. The parity-conserving part of the weak interaction induces, through the exchange of the $Z^0$ boson, a potential of
\beq\label{Vweak}
V_{\rm weak}(r)=-\frac{8G_{\rm F} m_{Z}^2}{\sqrt{2}}\frac{g_e g_A}{4\pi}\frac{e^{-m_Z r}}{r}\,,
\eeq
where $G_{\rm F}\approx1.166\times 10^{-5}\,$GeV$^{-2}$ is the Fermi constant, and $m_Z$, $g_e$ and $g_A$ are the mass and the vector couplings of the $Z^0$ boson to the electron and the nucleus, respectively. The tree level coupling values in the SM are $g_e^{\rm SM}=-1/4+s_W^2$ and $g_A^{\rm SM}=Q_W^{\rm SM}/4$, where $Q_W^{\rm SM}=-(A-Z)+Z(1-4s_W^2)$ is the SM tree level nuclear weak charge and $s_W^2\approx0.23$ is the sine of the weak mixing angle squared. The electron coupling was best probed at LEP through precision measurements at the $Z$ pole, and was found to agree (including radiative corrections) with the SM at the $10^{-3}$ level~\cite{Schael:2013ita}. On the other hand, deviation from the SM $Z^0$ couplings to the first generation quarks are poorly contrained in a model-independent way. In particular, new physics contributions to the right-handed up and down quark couplings  could be as large as $\cO(1)$ relative to their SM values~\cite{Efrati:2015eaa}. We therefore assume that the $Z^0$ coupling to electron is SM-like and use IS measurements to probe the parity conserving part of the first-generation quark couplings.
From Eq.~\eqref{Vweak}, the weak force contributes to the IS in Eq.~\eqref{IStheory} as 
\beq
X_{AA'}^i\big|_{\rm weak} = -1.3 \,{\rm Hz}\times q_W(A-A')\frac{|\psi(0)|^2}{4a_0^{-3}}\,,
\eeq
where $q_W$ is the nuclear weak charge per unit neutron, whose tree level SM value is $q_W^{\rm SM}=-1$. A deviation $\delta q_W\equiv q_W-q_W^{\rm SM}$ of the neutron weak charge can then be bound as follows. Substract first the weak contribution as predicted by the SM from the measured values of $\delta \nu^i_{AA'}$ and  construct a King plot with the residuals. A linear King plot in Yb$^+$ would then yield
\beq\label{Zbound}
|\delta q_W| \lesssim \frac{7.4\times 10^{-2}}{|1-F_{21}|}\left(\frac{\Delta}{{\rm Hz}}\right)\left(\frac{8}{A-A'}\right)\,,
\eeq          
Using $\delta q_W = 4\delta g_u+8\delta g_d$, where $\delta g_{u,d} \equiv g_{u,d}-g_{u,d}^{\rm SM}$ with $g_u^{\rm SM} = 1/4-2s_W^2/3$ and $g_d^{\rm SM}=-1/4+s_W^2/3$ at tree level,
 the above constraint translates into bounds on deviations from the SM of the $Z^0$ coupling to fundamental quarks as
\beq\label{Zbound}
|\delta g_{u}+2\delta g_d|\lesssim 1.8\times10^{-2}\,. 
\eeq
Although Eq.~\eqref{Zbound} applies to the parity-conserving parts only, it is  stronger than model-independent bounds~\cite{Efrati:2015eaa} from LEP by a factor $\sim2$ for up-quarks, and about an order of magnitude for  down-quarks.
Equation~\eqref{Zbound} shall be compared to a similar bound obtained from atomic parity violation measurements in $^{133}$Cs atom~\cite{wood1997measurement,guena2005measurement}. Due to the accidentally small $Z$ coupling to protons, APV is mostly sensitive to $q_W$,  giving $|\delta g_u+2\delta g_d|\lesssim 10^{-3}$~\cite{PDG}. Albeit weaker, the IS constraint is complementary to the latter since it is insensitive to the presence of additional sources of parity-violation beyond the weak force.

\section{New Physics below the GeV scale} \label{sec:LNP}

New physics mediators $\phi$ with mass $m_\phi$ below the GeV scale are not captured by the EFT analysis above, since the range of the new force is parametrically larger than the nuclear size. The potential between nuclei and their bound electrons is parameterized generically as
\begin{align}
	\label{eq:Vlight}
	V_{\rm light} (r) =(-1)^{s+1} \alpha_\phi N_e N_A\frac{e^{-m_\phi r}}{r} \, ,
\end{align}
where $s=0,1,2$ is the spin of the mediator, $\alpha_\phi$  its coupling strength and $N_e$ and $N_A$ denote respectively the charge of the electron and the nucleus under the new $\phi$-mediated force. In contrast with the previous cases, the shift $\delta E_k$ of an energy-level $k$ now requires the knowledge of the electron wave function in the entire atomic volume, and not just $|\psi(0)|^2$.  In heavy atoms or ions,  the shape  of the wave function at all radii is strongly sensitive to electron-electron correlation effects and relativistic corrections. Many-body simulations are necessary to accurately account for these effects, which is far beyond the scope of this work. 
\begin{figure}[!t]
\includegraphics[width=0.4\textwidth]{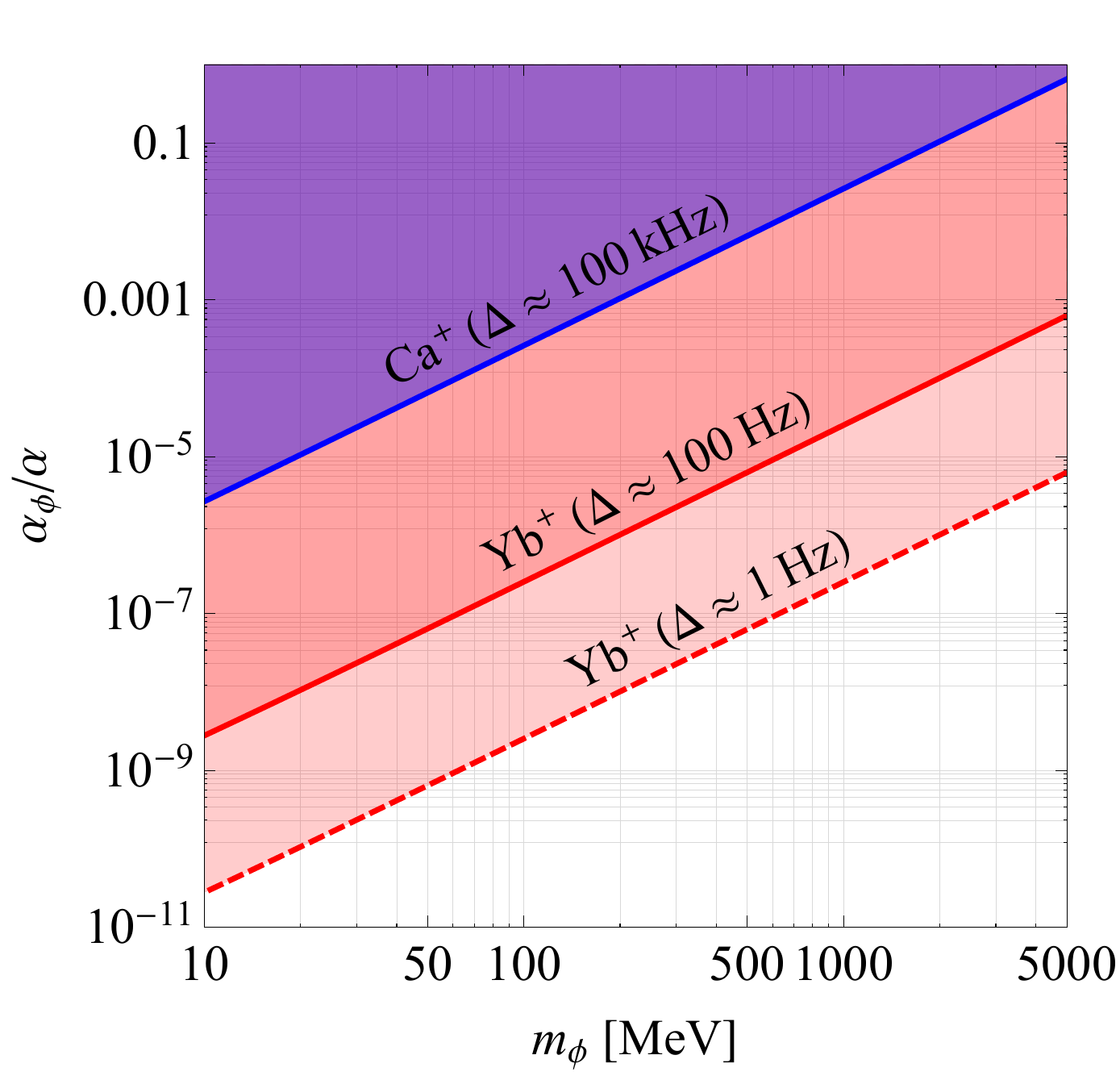}
\caption{Regions of the mass vs. coupling strength plane probed by IS measurements in Ca$^+$~\cite{isotope_shift_measurement} (purple) and Yb$^+$ (red), for sub-GeV mediators. For Yb$^+$, an experimental accuracy of $\Delta=\,$Hz (dashed) and $\Delta=100\,$Hz (solid), as well as $A-A'=8$ and $|1-F_{21}|\simeq1$ are assumed.}
\label{fig:light}
\end{figure}
However, for $r\lesssim a_0/Z$, the screening of the other electrons is negligible and the wave function  in the inner part of the atom is well approximated by one in the non-relativistic limit~\cite{AtomicBook}  
\begin{align}
	\label{eq:Psie}
	\psi(r\lesssim a_0/Z) \simeq \psi(0) e^{-Zr/a_0 } \, .
\end{align}
The above approximation is expected to break for interaction ranges $1/m_\phi$ longer than $a_0/Z$ or, equivalently, for mediator masses satisfying $m_\phi\lesssim Z \alpha m_e\sim 100\,$keV-$1\,$MeV for $Z=70$. 
The resulting contribution to the IS in Eq.~\eqref{IStheory} is found to be
\beq
X_{AA'}^i\big|_{\rm light} &\simeq& 5\times 10^7 \,{\rm Hz}\times (-1)^{s}\alpha_\phi N_e(N_A-N_{A'})\nonumber\\
&&\times \frac{|\psi(0)|^2}{4a_0^{-3}}\left(\frac{{\rm GeV}}{m_\phi+2Za_0^{-1}}\right)^2\,.
\eeq
Note that if $\phi$ couples only to  protons in the nucleus, then $N_A-N_{A'}=0$ and the IS contribution vanishes. For sake of concreteness, we assume in the rest of this section $N_e=1$ and, in analogy with the Higgs force~\cite{Delaunay:2016brc}, $N_A=A$. We show in Fig.~\ref{fig:light} the values of $\alpha_\phi$ probed with IS measurements for mediator masses in the range $10\,$MeV$<m_\phi< 10\,$GeV. Existing Ca$^+$ data~\cite{isotope_shift_measurement} constrain couplings of $\mathcal{O}(10^{-6}\alpha)$ for $m_\phi=10\,$MeV, while King plots constructed from prospective Yb$^+$ measurements at the Hz level could improve these bound by five orders of magnitude. 

\section{On a possible $750\,$GeV resonance} \label{sec:X750}

%
\begin{table}[!t]
\begin{center}
\renewcommand{\arraystretch}{1.3}
\begin{tabular}{|c|c|c|}
\hline
 $S$ couplings & LHC\,(8,\,13) bound~\cite{Aad:2014cka,ATLAS-CONF-2015-070}  &  IS projection \\
 ($\mu=750\,$GeV) & ($\Gamma_S=45\,$GeV) &  ($\Delta=1\,$Hz)\\
 \hline\hline
  $|y_ey_u|$ & $(5.6,\,6.0)\times 10^{-3}$ & $1.5\times 10^{-3}$ \\
  $|y_ey_d|$ & $(7.3,\,7.8)\times 10^{-3}$& $1.2\times 10^{-3}$\\
  $|y_ey_s|$ & $(2.9,\,2.5)\times 10^{-2}$& $1.5\times 10^{-2}$\\
\hline
    $|y_ey_c|$ & $(3.6,\,3.0)\times 10^{-2}$ & $9.6\times 10^{-2}$\\
  $|y_ey_b|$ & $(5.6,\,4.5)\times 10^{-2}$ & $0.49$\\
  $|y_ey_t|$ & $(0.19,\, 0.16)$ & $32$\\
\hline
    $|y_ec_g|$ & $(0.72,\,0.60)$ & $150$\\
  \hline
  \hline
\end{tabular}
\end{center}
\caption{Bounds on a 750$\,$GeV scalar resonance couplings from direct searches in $e^+e^-$ final states at the $8\,$TeV and $13\,$TeV LHC  and from prospective isotope shift measurements in Yb$^+$ clock transitions. The LHC bounds assume a resonance width of $\Gamma_S=45\,$GeV and scale like $\sqrt{\Gamma_S}$. The IS projected bounds assume an experimental accuracy of $\Delta=1\,$Hz and $A-A'=8$, and scale like $\Delta/(A-A')$. }
\label{tab:750}
\end{table}%

We move now to consider a resonance $S$ with a mass of $750\,$GeV, possibly related to the diphoton excess recently reported by the ATLAS and CMS experiments at the 13$\,$TeV LHC~\cite{ATLAS-CONF-2015-081,CMS-PAS-EXO-15-004}. This observation corresponds to an estimated $pp\to S\to \gamma\gamma$ cross section of $\sigma^{13}_{\gamma\gamma}\approx 5\,$fb~(see {\it e.g.}~\cite{Franceschini:2015kwy}). Furthermore, ATLAS result weakly favors a large resonance width of $\Gamma_S\approx45\,$GeV. Production of this new physics state at the LHC already establishes its coupling to protons, either through gluons, valence or sea quarks. From the observation of a prompt decay into diphotons, we know that $S$ is either a scalar, pseudoscalar or a spin two particle~\cite{Landau,Yang}. If $S$ is a parity even state which further couples to electrons and neutrons, there is an opportunity to probe its coupling structure with IS measurements. For illustration we will focus
on the scalar case, however our results are straightforward to generalize for a spin two.  
The relevant $S$ couplings to SM states are parameterized by the following phenomenological Lagrangian at the scale $m_S=750\,$GeV
\beq
-\cL_{S} = y_f S \bar f f -\frac{\alpha_sc_g}{12\pi m_S}S(G_{\mu\nu}^a)^2- \frac{\alpha c_\gamma}{2\pi m_S} S F_{\mu\nu}^2\,,
\eeq 
where $f=e,q,Q$ with $q=u,d,s$, $Q=c,b,t$ and $\alpha_s$ is the QCD coupling. For energy scales $E\leq m_S$, the relevant effects of $S$ are parameterized at leading order by the effective operators in Eqs.~\eqref{OpeqS}--\eqref{OpeQS} with $\Lambda=m_S$ and  
$c_{eq}^S=y_e y_q$, $c_{eQ}^S=y_e y_Q$, $\delta c_{eg}= -y_ec_g/(12\pi)$,  
while $c_{eq}^V=0$. At the GeV scale, the $c,b,t$ quarks are integrated out and contribute as threshold corrections to $\cO_{eg}$ whose WC is given by Eq.~\eqref{thresholdcorr} with $\Lambda=m_S$.
In order to derive the constraints on  $S$ couplings to fermions and gluons from Yb$^+$ measurements, we simply recast the projected bounds in Table~\ref{tab:EFT}  
from the EFT analysis. 
Similar combinations of $S$ couplings are bounded by direct resonance searches decaying into $e^+e^-$ at the LHC. The best bounds on a $pp\to S\to e^+e^-$ resonance are set by LHC data at $8\,$TeV and $13\,$TeV with a cross section of $\sigma_{ee}^8\lesssim 2\,$fb~\cite{Aad:2014cka} and $\sigma_{ee}^{13}\lesssim 6.5\,$fb~\cite{ATLAS-CONF-2015-070} at $95\%$ confidence level (CL), respectively. 
Assuming that $S$ production at the LHC is dominated by either $c_g$ or one of the quark couplings yields conservative direct bounds on the product of $y_e$ and $y_q$, $y_Q$ and $c_g$~\cite{Franceschini:2015kwy,Gupta:2015zzs}. We summarize these bounds together with the projected ones from IS measurements in Table.~\ref{tab:750}.
We learn that IS measurements with state-of-the-art accuracy in Yb$^+$ could already surpass direct searches at the 8$\,$TeV and 13$\,$TeV LHC in probing $S$ couplings, unless the dominant production mechanism of the resonance is gluon fusion, through a loop of either top quarks or new physics states, or $c\bar c$ and $b\bar b$ annihilation.  

\begin{figure}[!t]
\includegraphics[width=0.47\textwidth]{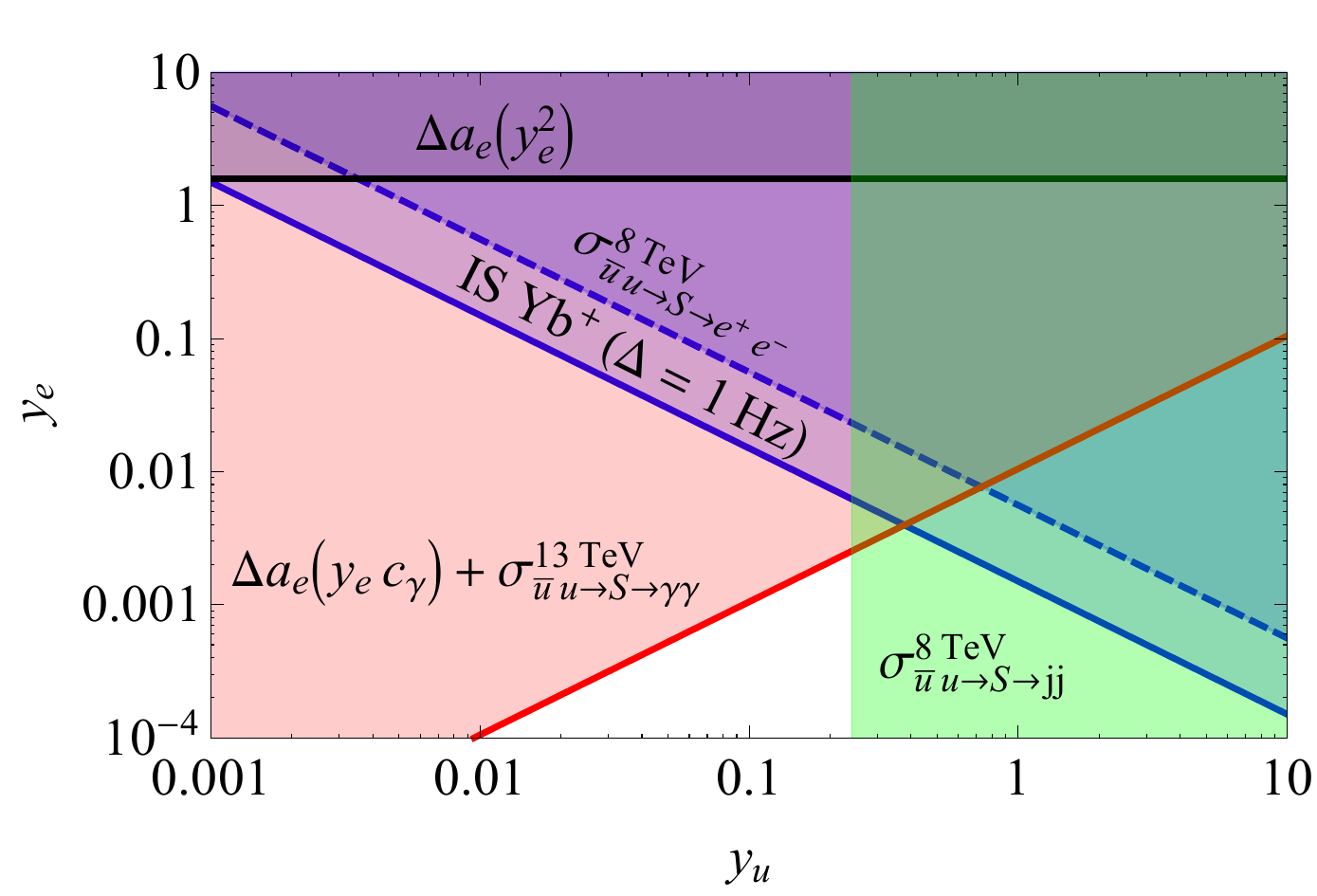}
\caption{Constraints on fermion couplings of a possible 750$\,$GeV resonance, assuming dominant LHC production from $u \bar u$ annihilation. Colored region are excluded, see text for details.}
\label{fig:boundsye750}
\end{figure}

The electron coupling is also indirectly constrained by precision measurements of  the anomalous magnetic moment of the electron $a_e \equiv (g_e-2)/2$~\cite{PhysRevLett.100.120801,Aoyama:2014sxa}, together with extracted values of $\alpha$ from other measurements in Rb atoms~\cite{bouchendira2011new}. As a result, new physics contributions to $a_e$ are severy constrained with $|\Delta a_e|\lesssim8.1\times 10^{-13}$ at 95$\%$\,CL~\cite{Giudice:2012ms,Altmannshofer:2015qra}. The dominant $S$ contribution to $g_e-2$ is through two one-loop diagrams scaling like $y^2_e$ and  $y_e c_\gamma$, respectively: $\Delta a_e\big|_{750} = y_e^2 F_{e} + y_e c_\gamma F_ {e\gamma}$~\cite{Goertz:2015nkp}. The first contribution is suppressed by $m_e^2/m_S^2$ and evaluates to $F_e\approx 3.1\times 10^{-13}$~\cite{Jackiw:1972jz}. On the other hand, the second diagram is only $m_e/m_S$ suppressed but is logarithmically sensitive to unknown UV physics.  It is nevertheless reasonably estimated through naive dimensional analysis~\cite{Manohar:1983md}, yielding $F_{e\gamma}\sim 2\times 10^{-11}$~\cite{Goertz:2015nkp} where an $\cO(1)$ uncertainty from the logarithm of the UV scale to $m_S$ ratio is understood. Barring accidental cancellations between the two diagrams, the $g_e-2$ constraint gives $|y_e|\lesssim 1.6$ and $|y_e c_\gamma|\lesssim 3.9\times 10^{-2}$, respectively. 
Combining the latter with a possible LHC diphoton signal yields an upper limit on the electron-to-production coupling ratio of  $|y_e|\lesssim(11y_u,8.3y_d,2.5y_s,2.1y_c,1.4y_b,0.11c_g)\times 10^{-3}$~\cite{Gupta:2015zzs}, which together with the bounds of Table~\ref{tab:750} allows to bound $|y_e|$ from above. Assuming the projected IS bounds gives $|y_e|\lesssim (4.1,3.2,6.1,14,26,130)\times 10^{-3}$. The above interplay is illustrated on Fig.~\ref{fig:boundsye750} in the case where $S$ production is $u\bar u$ dominated, along with LHC constraints on dijet production~\cite{CMS-PAS-EXO-14-005}.\\

\section{Outlook} \label{sec:Outlook}

In this letter we discussed the possibility to probe physics beyond the Standard Model with King plots constructed with precision Isotope Shifts~(IS) measurements in optical atomic clock transitions. Relying on an effective field theory approach, we derived the generic reach on new physics whose characteristic mass scale is well above $1\,$GeV. While bounds on spin-one mediated interactions are expected to be comparable to those from collider experiments, precision IS measurements could probe scalar mediators up to $\sim20\,$TeV if their neutron coupling is dominated by up or down quarks. In the case of sub-GeV forces, we estimated the sensitivity on the interaction strength to be $10^{-11}$ relative to the fine structure constant $\alpha\approx 1/137$ for mediator masses as low as 10$\,$MeV. Smaller masses could in principle be probed, provided an accurate knowledge of the electronic wave functions in the entire atomic volume which is beyond the scope of our preliminary analysis.
Furthermore, we proposed a simple way to probe deviations from Standard Model $Z^0$ couplings to up and down quarks. We found that prospective IS constraints are stronger than the LEP measurements at the $Z^0$ pole up to an order of magnitude for down quarks. Finally, we comment on the prospects to probe the coupling structure of a possible 750$\,$GeV scalar resonance. In particular, we showed how sensible knowledge on the resonance coupling to electrons can be extracted from IS measurements together with the LHC diphoton signal and indirect constraints from $g_e-2$.
 
Precision IS measurements open a new door to probe physics beyond the Standard Model. The complementarity of the informations gleaned with such low energy processes and those from the energy and intensity frontiers is a valuable asset towards improving our understanding of Nature's fundamental clockworks. 

\section*{Acknowledgments} \label{sec:Acks}
We thank Roee Ozeri and Jesse Thaler for comments on the manuscript.  
The work of CD is supported by the ``Investissements d'avenir, Labex ENIGMASS''.  The work of YS is supported by the U.S. Department of Energy under grant Contract Number  DE-SC0012567.

\bibliographystyle{apsrev}
\bibliography{BSM_Atomic_IS-bib}

\end{document}